\documentclass[noline]{copernicusArxiv}

\begin{document}

\graphicspath{{CASI2014/}}

\title{Joint Inference of Misaligned Irregular Time Series with Application to Greenland Ice Core Data}

\Author[1]{Thinh K.}{Doan}
\Author[2]{Andrew C.}{Parnell}
\Author[1]{John}{Haslett}

\affil[1]{School of Computer Science \& Statistics, Trinity College Dublin, Ireland}
\affil[2]{School of Mathematical Sciences, University College Dublin, Ireland}


\runningtitle{Joint inference of misaligned time series}
\runningauthor{Doan et al.}
\correspondence{T. K. Doan (tdoan@tcd.ie)}



\firstpage{1}

\maketitle

\begin{abstract}
Ice cores provide insight into the past climate over many millennia. Due to ice compaction, the raw data for any single core are irregular in time. Multiple cores have different irregularities; jointly these series are misaligned. After processing, such data are made available to researchers as regular time series: a data product. Typically, these cores are independently processed. In this paper, we consider a fast Bayesian method for the joint processing of multiple irregular series. This is shown to be more efficient. Further, our approach permits a realistic modelling of the impact of the multiple sources of uncertainty. The methodology is illustrated with the analysis of a pair of ice cores (GISP2 and GRIP). Our data products, in the form of marginal posterior distributions on an arbitrary temporal grid, are finite Gaussian mixtures. We can also produce sample paths from the joint posterior distribution to study non-linear functionals of interest. More generally, the concept of joint analysis via hierarchical Gaussian process model can be widely extended as the models used can be viewed within the larger context of continuous space-time processes.\end{abstract}

\introduction  
Ice cores play an important role in revealing Earth's climate history via the analysis of their chemical composition. Data from ice cores are typically available in two forms: "raw" (irregular in time) and "data products" (regularly-spaced in time, i.e. "gridded"). Often the latter are used as input for climate models to analyse past climate change. Alternatively, being gridded, researchers can simply combine one series with other data series, similarly gridded, representing other aspects of climate. Such data products are pre-processed from raw data using a variety of techniques; from simple running averages \citep{stuiver2000gisp2, thomas20078} to complex parametric smoothing \citep{peavoy2010bayesian, nieto2014bayesian}. These imputation techniques - broadly interpolators or smoothers - are typically applied one core at a time. 

This paper proposes a joint statistical model for processing multiple misaligned ice core time series data. Like others in the wider field of palaeoclimate reconstruction \citep[e.g.][]{haslett2006bayesian, tingley2012piecing}, our procedure is based on hierarchical Gaussian processes approached through Bayesian inference. Our use of a joint approach is more efficient than its independent alternative as it permits a sharing of strengths across multiple cores, with each core complementing the other in the information needed to learn about the true underlying latent process. A pair of ice cores drilled in Greenland are used to illustrate our proposed framework. Our primary goal is to efficiently take into account multiple sources of information to produce interpretable imputations. A secondary goal is to encourage the use of sample paths of such processes as a richer  type of "product". 

We introduce and discuss the raw data and their associated uncertainties in Sect. \ref{sec:ice}. In Sect. \ref{sec:methods} we describe our hierarchical Bayesian stochastic process model, and discuss an efficient procedure for inference of the underlying latent process. We report and discuss some results in Sect. \ref{sec:results}. One of the advantages of our approach is that researchers can use process sample paths to study non-linear functionals of partially observed processes. As typically published, classical interpolants and smoothers do not facilitate such studies, for they do not typically make available the joint uncertainty associated with the product, a point made pithily by \citet{mcShane2011statistical} in their critical study  of surface temperature reconstructions in the last millennium. Details of the so-called "8.2ka event" \citep[e.g.]{thomas20078}, an abrupt climate change period, provide specific examples. For instance, what precision do such data permit in the timing of this event? In Sect. \ref{sec:82k}, we demonstrate that sample paths from our model are useful to answer this question. In Sect. \ref{sec:comparisons}, we compare various aspects of the performance of our proposed model against alternative specifications.


Whilst our motivating example stems from climate research, multivariate data with different temporal irregularities are a common feature in many contemporary applications. For example, the ability to combine outputs at different levels of accuracy is crucial to the understanding of processes being studied through potentially expensive computer experimentation. A useful approach in such applications is to combine results from many, cheap but low-resolution experiments with those from a few, expensive but high-resolution experiments, by linking the data via another layer of modelling \citep{qian2008bayesian}. In medical applications, one recommendation to overcome issues with misalignment is to align the times before further modelling \citep{cismondi2013missing}. For more examples of misaligned time series and associated methods, see \citet{cismondi2011computational} and \citet{eckner2012framework} and the references therein. 

From a theoretical perspective, we view misaligned time series as a special case of spatial misalignment in spatial statistics \citep{gelfand2001change, wikle2005combining}. More generally, our methodological proposal follows the "change of support problem" for data that are indexed in both space and time. This is a generic term that describes methods to combine multiple data sets and make prediction at scales or resolutions which may be different to those of the observations. In this work, we focus on the case where the misaligned observations are at points in time and seek prediction of underlying processes at arbitrary new time points,  typically on a regular grid. We offer our perspectives on possible directions of future theoretical and empirical research in Sect. \ref{sec:discussion}. 


\section{Greenland ice core data}\label{sec:ice}

Climate indicators such as tree rings, laminated lake sediments and ice cores are often used as a guide to past climatic conditions \citep{tingley2012piecing}. This paper focuses on the stable isotope ratios of oxygen in ice cores that are linked to past temperature. The process by which an ice core represents temperature is based on the ease with which two particular types of isotopes in water evaporate from the ocean and condense as snow. We refer interested readers to \citet{jouzel1997validity} for a more detailed account of this connection. The measurement is $\delta^{18}O$, recorded as a function of depth. At least for the Holocene period considered in our work, the depth information is transformed to age by counting annual layers in chemicals that show a seasonal cycle \citep{rasmussen2006new}. 

We obtain data from the National Snow and Ice Data Center, University of Colorado at Boulder, and the World Data Center for Paleoclimatology, National Geophysical Data Center, Boulder, Colorado (\url{http://www.ncdc.noaa.gov/paleo/icecore/greenland/}). In particular, we use the datasets from the United State's Greenland Ice Sheet Project Two \citep[GISP2]{stuiver1999gisp2} and Europe's Greenland Ice Core Project programs \citep[GRIP]{johnsen1999grip}. These are the results from drilling through the Greenland Ice Sheet to recover ice records over 3 km deep. The data sets are presented in several versions; we use a raw dataset in which the cores were cut into sections of equal lengths; 55cm and 200cm for GRIP and GISP2 respectively. A data point is the averaged $\delta^{18}O$ measurement defined at a time point centred at each section of a core. Hence each section of the core can be viewed as the support of a measurement. Even though the sections are equal in length, the ages are irregular. Thus, jointly, multiple cores with different irregularities are temporally misaligned. 

We consider the Holocene, a period of relatively stable climate between 0 - 11.5 thousand calendar year Before Present (k cal yr BP), where Present is 1950. $\delta^{18}O$ and date records for all consecutive sections are presented in Fig. \ref{fig:input1}, along with boxplots of the age increments. Note that we have omitted an age difference value of 80.6 yr between roughly 1320 and 1400 cal yr BP to focus on other significant features of this plot. We see different irregularities in the age increments. Moreover, we note that the ratio of their median values is roughly the same as the ratio in the lengths of the sections, i.e. $\frac{55}{200}$. We discuss aspects of this in App. \ref{app:cosp}.

The empirical semivariograms in Fig. \ref{fig:cosemp} suggest that each series may be adequately modelled via a linear variogram. We defer the discussion of variogram modelling to Sect. \ref{sec:fullmodels}. For the purpose of exploration, we use the R package geoR \citep{ribeiro2001geor} to estimate the parameters of the semivariogram models using the weighted least square method, and compute the total variability of each increment of the $\delta^{18}O$ measurements. We then create QQ plots of the standardised samples as presented in Fig. \ref{fig:input2}. Apart from two unusual values from each core, these plots indicate the suitability of the Gaussian assumption for the increments. Upon further investigation, we found that the unusual values correspond to neighbouring pairs of measurements with very large negative difference in the $\delta^{18}O$ values.



\section{Models and inference}\label{sec:methods}
In this section we outline our notation and describe our model for misaligned time series data. Subsequently we show how to perform fast inference on this model without resort to Markov chain Monte Carlo methods, and describe an algorithm for efficient imputation of the latent process onto a grid.


\subsection{Hierarchical model}\label{sec:fullmodels}

To simplify the narrative in the rest of this article we introduce some notation. Let $y(t_{c,i}); i = 1...n_c$ denote the observed values of $\delta^{18}O$ in core $c$ at time $t_i$; as  a vector, we write $\boldsymbol{y}_c$. We label the cores as $c = 1, \ldots m$. There are $N = \sum_c n_c$ observations in total, and we use $\boldsymbol{y}$ to denote these. 

We consider a hierarchical model comprised of two main layers. At the observation level $z(t_{c,i})$ is an unobserved value of $\delta^{18}O$. When it is observed, the instrumentation is such that 
\begin{equation}
	y(t_{c,i}) = z(t_{c,i}) + \nu(t_{c,i})
\end{equation} 
where the terms $\nu(t_{c,i})$ are iid, zero mean, Gaussian random variables with fixed known variance $\sigma^2_{\nu_c}$ corresponding to the instrumentation. For simplification, hereafter we drop the $i$ subscript when discussing latent processes since they are defined for all times. 

At a deeper level, we express $z(t_c)$ as a function of latent value $x(t_c)$ via an additive Gaussian model:
\begin{equation}\label{eq:data}
	z(t_c) = x(t_c) + w(t_c)
\end{equation} 
We propose to model $x(t_c)$ as a continuous-time, independent increments process with increment variance such that for each core $Var\left[x(t_c) - x(t_c - h)  \right]  = v^2 |h|$, i.e. having a linear variogram with no nugget effect. This process may be thought of as driven by climate and thus slowly changing. The process $w(t_c)$ reflects micro-scale, annual-level variations and is modelled by a white noise process, independently across cores, having common $Var\left[ w(t_c) \right]  = \sigma^2 _{w_c}$.  Consequently we have
\begin{equation}\label{eq:alltogether}
	y(t_c) = x(t_c) + \nu(t_c) + w(t_c) := x(t_c) + \epsilon(t_c)
\end{equation} 

Here \(\epsilon(t_c) \overset{ind}{\sim} \mathcal{N}(0, \sigma^2_{\epsilon_c})\), with $\sigma^2_{\epsilon_c} =  \sigma^2_{w_c} + \sigma_{\nu_c}^2$, where the core-specific variance component $\sigma^2_{\epsilon_c}$ is the annual-level nugget effect at core $c$. The term ''nugget'' here refers to both sources of variation: the noise at time differences much smaller than the sampling interval, and that which is due to uncertainty from data collection \citep[e.g.,][Chap. 2.4]{chiles2012geostatistics}.

In Sect. \ref{sec:ice}, we argue that the linear variogram is a suitable model for both ice core data sets. More formally each is consistent with a generating process which is the sum of: (i) an underlying independent increments continuous time process with variogram proportional to lag - the slope; and (ii) a white noise process, manifest in the intercept or nugget effect which is dominated by $\sigma^2_{w_c}$. For our application, the measurement processes differ for each core. In particular both involve averaging a section of core. The chief implication of this is that the nugget effects differ in proportion to the time durations associated with the sections. We can reparameterize so that there is only one nugget term $\sigma^2_\epsilon$ for one of the cores, with the other having a different nugget which is a multiple of this value, i.e., $\sigma^2_{\epsilon_c} = k_c \sigma^2_\epsilon$ for a positive known core-specific value $k_c$ based on the ratio of the measurement periods. For GRIP and GISP2, we set this constant to be the ratio of average supports of 0.275. We outline this choice in more detail in App. \ref{app:cosp}. 

The $x$-increments at different cores will be highly correlated if they reflect climate, and are physically close. In vector forms, $\boldsymbol{x}$ is the multivariate latent process for all cores at all times, and $\boldsymbol{x}(t)$ represent all the cores at time $t$, we write
\begin{equation}\label{eq:state}
\boldsymbol{x}(t + h) - \boldsymbol{x}(t) ~\sim~ \mathcal{N}(\boldsymbol{0}, v^2~|h|~\boldsymbol\Sigma)
\end{equation}  
where $v^2$ is the common marginal variance of an increment per unit time, and $\mathbf\Sigma$ is an $m$-by-$m$ matrix which controls the strength of the relationship of data across the cores. The multivariate process in Eq. \eqref{eq:state} forms the basic model underlying our joint approach. To complete the hierarchical specification, prior distributions are assigned to the hyperparameters $\boldsymbol{\theta}=(v^2, \sigma^2_{\epsilon_c}, \Sigma)$. We use reference priors on $v^2$, $\sigma^2_{\epsilon_c}$ so that, for example, $\pi(v^2) \propto v^{-2}$. We defer the discussion of model choice for $\mathbf\Sigma$ to Sect. \ref{sec:comparisons}.

\subsection{Imputation}\label{sec:impute}

Our objective, given $\boldsymbol{y}$, is to provide imputed values of $\delta^{18}O$ for all cores on a regular time grid denoted by $\boldsymbol{t_g} = \{i \Delta; i=1, \ldots, n_g\}$. These values will be proposed below as expected values of suitable random variables, conditional jointly on $\boldsymbol{y}$. Our contribution in this paper is to focus on joint conditioning, in contrast to conditioning on data $\boldsymbol{y}_c$ in a single core, and in the issues of uncertainty that arise when the various parameters above are themselves only available through statistical inference. 

There are two choices of random variable for imputation. We could focus on $E[z(t_c) | \boldsymbol{y}]$ or $ E[x(t_c) | \boldsymbol{y}]$ for each core $c$ and times $t \in \boldsymbol{t_g}$. Both are legitimate targets of interest; and their computations are equally straightforward. In this paper we choose the latter; see Eq. \eqref{eq:cGmean}. Given the parameters appropriate to our data, and in particular that $\sigma^2_{w_c} \gg \sigma^2_{\nu_c}$ there will be little difference between them, except for cases where a time grid effectively coincides with a time of an observation, and for the corresponding core only.  The associated variances,  $Var[z(t_c) | \boldsymbol{y}]$ and $Var[x(t_c) | \boldsymbol{y}]$, will however differ; conditional on the parameters this difference will be approximately $\sigma^2_{w_c}$. We return to this in Sect. \ref{sec:results}.


\subsection{Inference}\label{sec:postHyper}


Our later mathematical derivations are substantially simplified by defining: 
\begin{enumerate}
\item $\boldsymbol{t_o}$ to be the sorted set of all observed times at all cores, i.e. $t_o = \{t_{c,1}, t_{c,2}, \hdots\} = \{t_1, t_2, \hdots, t_n\}$ where $n$ is the total number of unique times across all cores. 
\item ${{y}_{o}(t_{c,i})}; c=1, \ldots, m$  whose values coincide with the ${y}(t_{c,i})$ values when core $c$ has an observation at $t_i$ and whose values are missing otherwise; there is typically one such core for each $t_i$. Let $\boldsymbol{y_o}$ denote the set of all such vectors.
\end{enumerate}
This is only a notational trick; the vectors $\boldsymbol{y}$ (length $N$) and $\boldsymbol{y_o}$ (length $mn$) contain the same information. The objective is to use this information, together with a hierarchical model of the continuous time multivariate stochastic process $\boldsymbol{x}$ to impute its values on $\boldsymbol{t_g}$. We shall refer to $\boldsymbol{y}$ and $\boldsymbol{x}$ defined on $\boldsymbol{t_g}$ as $\boldsymbol{y_g}$ and $\boldsymbol{x_g}$. Moreover, we let $\boldsymbol{x_o}$ be the latent multivariate process $\boldsymbol{x}$ defined on $\boldsymbol{t_o}$.

\subsubsection{Posterior distributions}
 In our new notation we can rewrite some of the equations discussed in Sect. \ref{sec:fullmodels} in their compact matrix form. For instance Eq. \eqref{eq:alltogether} can be written as \( \boldsymbol{y_o} | \boldsymbol{x_o}, \sigma^2_{\epsilon} \sim \mathcal{N}(\boldsymbol{x_o}, \mathbf{Q}^{-1}_{\epsilon_o})\) where $\mathbf{Q}_{\epsilon_o}$ is a diagonal precision matrix with entries $(\sigma^{2}_\epsilon)^{-1}$ and $(k_c \sigma^{2}_\epsilon)^{-1}$ corresponding to cores at times with data, and zeros where there are no data. This is the notation trick we previously introduced to allow us to write all vectors and matrices in clear order of time and core identity. Note that no computation is ever necessary for components that do not represent data. Similarly, Eq. \eqref{eq:state} has the form \( \boldsymbol{x_o} | v^2, \boldsymbol\Sigma ~\sim~  \mathcal{N}(\boldsymbol{0}, \mathbf{Q}^{-1}_{x_o}) \) where $\mathbf{Q}_{x_o} = \mathbf{Q}_{t_o} \otimes {v^{-2}\boldsymbol\Sigma}^{-1}$. Here, $\mathbf{Q}_{x_o}$ is a Kronecker product of $\mathbf{Q}_{t_o}$, the precision matrix of the univariate independent increment process for irregularly-spaced time series data \citep[Sect. 3.3]{rue2005gaussian}, and correlation matrix $\mathbf\Sigma$. We can now write the full posterior distribution as 
\begin{align}
\pi(\boldsymbol{x_o}, \boldsymbol\theta | \boldsymbol{y_o}) &\propto ~ \pi(\boldsymbol{y_o} | \boldsymbol{x_o}, \boldsymbol\theta) \pi(\boldsymbol{x_o} | \boldsymbol\theta) \pi(\boldsymbol\theta) \label{eq:hyperPost} \\
&\propto ~ |\mathbf{Q}_{\epsilon_o}|^\frac{1}{2}\exp\left(-\frac{1}{2}(\boldsymbol{y_o} - \boldsymbol{x_o})^T\mathbf{Q}_{\epsilon_o}(\boldsymbol{y_o}-\boldsymbol{x_o})\right) \nonumber \\
&\times \left|\mathbf{Q}_{x_o}\right|^\frac{1}{2} \exp\left(-\frac{1}{2}{\boldsymbol{x_o}}^T\mathbf{Q}_{x_o}{\boldsymbol{x_o}}\right)\pi(\boldsymbol\theta) \nonumber \\
&\propto ~ \mathcal{N}\left(\boldsymbol{x_o}; {\boldsymbol\mu}_{x_o|y_o}, \mathbf{Q}^{-1}_{x_o|y_o}\right) \left|\mathbf{Q}_{x_o|y_o}\right|^{-\frac{1}{2}} |\mathbf{Q}_{\epsilon_o}|^\frac{1}{2} \nonumber \\
&\times \left|\mathbf{Q}_{x_o}\right|^\frac{1}{2} \exp\left(\frac{1}{2}{\boldsymbol{y_o}}^T \mathbf{Q}_{\epsilon_o}(\boldsymbol\mu_{x_o|y_o} - \boldsymbol{y_o})\right)\pi(\boldsymbol\theta) \nonumber
\end{align}
Here \(\left|\mathbf{Q}_{x_o}\right| = \left|v^2 \mathbf\Sigma\right|^{-(n-1)} \), $\mathbf{Q}_{x_o|y_o} = \mathbf{Q}_{\epsilon_o} + \mathbf{Q}_{x_o}$, and  ${\boldsymbol\mu}_{x_o|y_o}$ is the solution to $\mathbf{Q}_{x_o|y_o}{\boldsymbol\mu}_{x_o|y_o} = \mathbf{Q}_{\epsilon_o} \boldsymbol{y_o}$. Thus $\boldsymbol{x_o}$ can be analytically integrated out of $\pi(\boldsymbol{x_o}, \boldsymbol\theta | \boldsymbol{y_o})$, yielding $\pi(\boldsymbol\theta | \boldsymbol{y}_o)$, before further computation. Thus, our inference procedure comprises two separate stages as follows.

\subsubsection{Inference stage 1: hyperparameters}\label{sec:inferstage1}
Initially, we focus on $\pi(\boldsymbol\theta | \boldsymbol{y_o})$, which the same as $\pi(\boldsymbol\theta | \boldsymbol{y})$. We begin with an optimization routine on the logarithm of $\pi(\boldsymbol\theta | \boldsymbol{y_o})$ to locate the mode $\hat{\boldsymbol\theta}$ and the Hessian matrix evaluated at $\hat{\boldsymbol\theta}$; the latter is asymptotically the precision matrix for $\hat{\boldsymbol\theta}$. These information will be used as a guide to explore the parameter space of $\pi(\boldsymbol\theta | \boldsymbol{y_o})$ via a search strategy proposed by \citet[Sect. 3.1]{rue2009approximate}. We approximate the continuous distribution for $\boldsymbol\theta$, namely $\pi(\boldsymbol\theta); \boldsymbol\theta \in \boldsymbol\Theta$ by the discrete distribution $\pi(\boldsymbol\theta_j); \boldsymbol\theta_j \in \boldsymbol\Theta_J;j=1, \ldots, J$. $\boldsymbol\Theta$ will be used as a numerical approximation to analytical integrations. For instance, the normalising constant can be discretely evaluated when there are few hyperparameters, i.e. \(
\pi(\boldsymbol{y_o}) = \int\limits_{\boldsymbol\theta  \in \boldsymbol\Theta} \pi(\boldsymbol{y_o}|\boldsymbol\theta) \pi(\boldsymbol\theta)d\boldsymbol\theta \approx \sum\limits_{\boldsymbol\theta _j \in \boldsymbol\Theta_J} \pi(\boldsymbol{y_o}|\boldsymbol\theta_j) \pi(\boldsymbol\theta_j)\triangle\boldsymbol\theta_j
\)

\subsubsection{Inference stage 2: imputation}\label{sec:cG}

Our next goal is to derive the marginal distribution of $\boldsymbol{x_g}$ given all observations. We repeat the aforementioned notational trick by letting the star notation denote the processes defined at both the unique (and sorted) observed times and grid, i.e. \(\boldsymbol{y_*} = \left({\boldsymbol{y_o}}^T,  {\boldsymbol{y_g}}^T\right)^T\) and \(\boldsymbol{x_*} = \left({\boldsymbol{x_o}}^T, {\boldsymbol{x_g}}^T\right)^T\). The problem becomes that of evaluating
\begin{equation}\label{eq:int}
\pi(\boldsymbol{x_*} | \boldsymbol{y_*}) = \int\limits_{\boldsymbol\Theta} \pi(\boldsymbol{x_*} | \boldsymbol{y_*}, \boldsymbol\theta) \; \pi(\boldsymbol\theta | \boldsymbol{y_*}) \; d\boldsymbol\theta \end{equation}

Derivation of the first quantity in the above integrand is, again, by completing the quadratic form as in Eq. \eqref{eq:alltogether}. The second quantity is effectively $\pi(\boldsymbol\theta | \boldsymbol{y})$, the joint posterior distribution of the hyperparameters previously derived in Sect. \ref{sec:inferstage1}. Importantly, the discrete approximation of the latter renders as summations the integrals that arise in Eq. \ref{eq:int}, i.e. 
\begin{equation}\label{eq:cGmarg}
\pi(\boldsymbol{x_*} | \boldsymbol{y_*}) \approx  \sum\limits_{\boldsymbol\Theta_J} \mathcal{N}\left(\boldsymbol{x_*}; \boldsymbol\mu_{x_* |y_*}(\boldsymbol\theta_j), \mathbf{Q}^{-1}_{x_*| y_*}(\boldsymbol\theta_j)\right) \pi(\boldsymbol\theta_j |\boldsymbol{y_*}) \triangle\boldsymbol\theta_j
\end{equation}
where $\boldsymbol\mu_{x_* |y_*}$ and $\mathbf{Q}_{x_*| y_*}$ are the posterior mean and precision of $(\boldsymbol{x_*} | \boldsymbol{y_*}, \boldsymbol\theta)$. We write $\boldsymbol\mu_{x_* |y_*}(\boldsymbol\theta_j)$ and $\mathbf{Q}_{x_*| y_*}(\boldsymbol\theta_j)$ to emphasize that they are the functions of $\boldsymbol\theta_j$. Thus, $\pi(\boldsymbol{x_*} | \boldsymbol{y_*})$ is a Gaussian mixture over the posterior samples $\boldsymbol\Theta_j$ with weights $\alpha_j = \pi(\boldsymbol\theta_j | \boldsymbol{y}) \triangle\boldsymbol\theta_j$ already computed in the first inference stage. The joint posterior distribution in Eq. \eqref{eq:cGmarg} will be useful for the simulation of sample paths. We return to this in Sect. \ref{sec:82k}.

The marginal posterior distribution of the $l^{th}$ element at a specific core corresponding to a temporal grid of interest may be approximated as finite Gaussian mixture:
\begin{equation}\label{eq:cGl}
\pi(x^{(l)}_* | \boldsymbol{y_*}) \approx  \sum\limits_{\boldsymbol\Theta_J} \mathcal{N} \left(x^{(l)}_*; \mu^{(l)}_{x_* |y_*}(\boldsymbol\theta_j), \tau^{(l)}_{x_*| y_*}(\boldsymbol\theta_j)\right) \alpha_j 
\end{equation}
where $\mu^{(l)}_{x_* | y_*}(\boldsymbol\theta_j)$, conditioning on the sample value $\boldsymbol\theta_j$, is the $l^{th}$ element of the corresponding core identity from $\boldsymbol\mu_{x_* | y_*}$. Similarly, each conditional posterior variance $\tau^{(l)}_{x_*|y_*}(\boldsymbol\theta_j)$ is the $l^{th}$ element of the core identity represented in the diagonal of the covariance matrix. In fact the latter can be computed efficiently from the precision matrix $\mathbf{Q}_{x_*| y_*}$ without having to perform matrix inversion \citep[Sect. 2]{rue2007approximate}. Such predictive distributions, in general non-Gaussian, are the source of any imputations proposed in Sect. \ref{sec:impute}. For example, an imputed value and its variance corresponding to the distribution in Eq. \eqref{eq:cGl} can be computed as
\begin{align}
E(x^{(l)}_* | \boldsymbol{y_*}) &\approx \sum\limits_{\boldsymbol\Theta_J} \mu^{(l)}_{x_* | y_*}(\boldsymbol\theta_j) \alpha_j \label{eq:cGmean} \\
Var(x^{(l)}_* | \boldsymbol{y_*}) &\approx \sum\limits_{\boldsymbol\Theta_J} \left(\mu^{(l)}_{x_* | y_*}(\boldsymbol\theta_j)\right)^2\alpha_j - \left(\sum\limits_{\boldsymbol\Theta_J} \mu^{(l)}_{x_* | y_*}(\boldsymbol\theta_j) \alpha_j \right)^2 \nonumber\\ 
&+ \sum\limits_{\boldsymbol\Theta_J} \tau^{(l)}_{x_*|y_*}(\boldsymbol\theta_j) \alpha_j \label{eq:cGvar}
\end{align}
Posterior means and variances are two summaries of the posterior marginal distributions. Calculations of other summarised statistics such as posterior modes, quantiles, etc. are also straight forward. Hence, for generality we use interquartile ranges (IQR) to quantify the imputations. We emphasize that although there are many computations with large (sparse) matrices, such as solving equations, Cholesky decomposition, etc. we only have to do this $J$ number of times. The fact that all of the matrices are sparse and can be efficiently stored and computed as band matrices presents a further huge reduction is the computation cost \citep{rue2005gaussian}.

\section{Model fitting results}\label{sec:results}

In this section we apply the model framework and inference procedures presented in Sect. \ref{sec:methods} to analyse the Greenland ice core datasets described in Sect. \ref{sec:ice}. When there are two cores, $m=2$. We model the cross-correlation matrix in Eq. \eqref{eq:state} as
\begin{equation}\label{eq:SigmaM1}
\boldsymbol\Sigma = 
\begin{pmatrix}
	1 & \rho \\
	\rho & 1
	\end{pmatrix}
\end{equation}
where $\rho$ is the correlation coefficient. Since temperature processes at nearby locations are always assumed to have (strongly) positively correlated increments, we set 0.5 and 1 as an acceptable range to be used in its prior. The hyperparameteres are \(\boldsymbol\theta = \{v^2, \rho, \sigma^2_\epsilon\}\), which represent the variance per unit increment of the latent process, cross correlation coefficient and nugget effect (for the GRIP core) respectively. Using the inference procedure outlined in Sect. \ref{sec:postHyper}, we obtain the discrete approximation to the marginal posterior distributions of $\boldsymbol\theta$. For brevity of presentation, we apply a smoother to the discrete distributions and present the smoothed version in Fig. \ref{fig:hyper}. It can be seen that the posterior distribution of $\rho$ peaks at a high value close to 1 which indicates a strong spatial relationship between the increments in two cores. This is not surprising since the physical locations of the cores are nearby and both reflect the historical changes in regional temperature of Greenland.

Our main interest lies in the data product derived from both cores jointly. The 95\% credible intervals for $\boldsymbol{x_g}$ at bidecadal time intervals are represented in Fig. \ref{fig:cG}.  We compare our results (labelled as DPH) with the data products of GISP2 and GRIP (labelled respectively as SG and R) as reported in \citet{stuiver2000gisp2} and \citet{rasmussen2006new}; both of which are based on averaging the measurements of shorter duration. The set of posterior imputed values from DPH can be seen to be much smoother than both SG and R because the nugget is so much larger. For both SG and R the nugget is assumed to be the measurement error which is set very small; see also \citep[Sect. 3.3]{bojarova2010non}. As we previously discussed in Sect. \ref{sec:fullmodels}, our nugget comprises both the measurement error and micro-scale variation. Our choice in Sec. \ref{sec:impute} removes the variation due to both sources of variation, resulting in a smoother latent process. 

To gain a better understanding of the benefit of joint modelling over independent alternatives, we fit an independent increments model with Gaussian noise to each core separately. Note that we suppress the relationship between two nugget effects as discussed in Sect. \ref{sec:fullmodels}; so that each model has two hyperparameters. We defer to Sect. \ref{sec:comparisons} for a more formal discussion of these separate models. In this section, we examine the IQR of posterior marginals of their latent process $\boldsymbol{x_g}$, in comparison to those of the joint model. It can be seen from Fig. \ref{img:indep} that the IQR in separate models for each core are always higher than those of the joint model. This result indicates that the joint approach utilises information from both cores more effectively when the relationship between the cores (here measured by $\rho$) is strong. We anticipate that the benefit of joint inference would be more apparent as the number of correlated cores increases.

Several interesting and informative features can also be seen in Fig. \ref{img:indep}. There is a consistently bigger difference in the IQR of the joint model and separate model for GISP2 than that for GRIP. This reflects the difference in the temporal resolution, or the number of available data points. The spikes (e.g. around times 0, 1.34, 8.2 k cal BP in GISP2) are also a direct implication of the gaps between times points. An exception is around the 8.2ka event, where the spikes are more influence by some abrupt changes in the $\delta^{18}O$ measurements. In general, the majority of the spikes in the IQR of the separate models are sharper than those of the joint model. However, the reverse happens around 1.4 k cal BP in GRIP. This is a direct consequence of the (lack of) data points from the other core (GISP2) at that period. Nevertheless, it remains consistent that this spike in the IQR of the joint model is lower than that from the separate model. Finally, we note that the IQR in all cores - in both the separate and joint setting - slightly increases with time. This occurs because temporal resolution decreases when the cores are sampled over sections of identical lengths. 

Although this is not a direct comparison with other methods - for neither standard deviations nor IQR are available - it suggests that these ignore valuable information by treating each core separately.

\section{Sample paths and the study of the 8.2ka event}\label{sec:82k}

The data product discussed in Sect. \ref{sec:results}, comprising gridded values, is conventional. It is arguably better than others - in the sense of reduced IQR by being based on all the data jointly. However, unaccompanied by appropriate measures of uncertainty, such products are of limited value as a basis for serious statistical research. As remarked forcefully by \citet{mcShane2011statistical}, even when accompanied by confidence intervals, as in Fig. \ref{fig:cG}, the value remains limited. For these are "\emph{pointwise} confidence intervals and are not confidence curves for the entire \emph{sample path}". That is to say, although the gridded values are jointly based on all the available observations, the joint conditional uncertainty of the unobserved $\delta^{18}O$ histories is not available. One manifestation is that the $\delta^{18}O$ history, as presented in a time series plot of the gridded values with a confidence band as in Fig. \ref{fig:cG}, is necessarily much smoother than the true (though unknown) history. One simple consequence is that uncertainties for changes in $\delta^{18}O$ are not available from such figures; for these minimally require covariances. The set of posterior means is in fact an average of all the possible histories that are statistically consistent with the model.

A greater challenge again is posed by 'events' such as the 8.2ka event, the sudden reduction in North Atlantic temperature during a period around 8.2 k cal yr BP \citep{thomas20078}. It is believed to be related to a transient change in the North Atlantic overturning circulation. Consequently, the amount of evaporated water in the ocean that became ice in Greenland is amongst its best sources of evidence. The date corresponding to the local minimum of the averages is not a satisfactory estimator of the time of such event. We use sample paths to illustrate a more satisfactory approach. Formally we focus on the random variables $x_{min} = min_t~x(t)$ and $t_{min} = argmin_t~x(t)$, being respectively the minimum value for each core, and the date on which this minimum was achieved.  Crucially these are non-linear functions of the latent process $x(t)$. Our objective is to make probabilistic statements about these, conditional on all the data. Sample paths provide the methodology.


For brevity, we use Tab. \ref{tab:82ka} to illustate an example of two sample paths from one core on a grid of 500 yr between time 7.5 and 9 k cal yr BP. These represent independent realisations of the latent process $\boldsymbol{x}$ on the time grid ($\boldsymbol{x_g}$). More specifically, they are simulated from $\pi(\boldsymbol{x_*} | \boldsymbol{y_*}, \boldsymbol{\theta})$, with hyperparameters $\boldsymbol{\theta}$ being a sample from $\pi(\boldsymbol\theta|\boldsymbol{y_*})$; both of which are  components of the integral \eqref{eq:int}. The non-linear functionals of interest can then be directly computed for each of very many sample paths and summarised.

A more complete version of this approach uses 1000 sample paths of the gridded process on a bidecadal time grid over the period of 7.90 to 8.50 k cal yr BP to study the 8.2ka event. Each corresponds  to a different fixed value of the sampled parameters $\boldsymbol\theta$, and conditional on  $\theta$ each sample path is of length 30 and is a draw from a Gaussian distribution. This time range is chosen to bracket the main event, and to distinguish from the possible long term climate trend \citep{morrill2005widespread}. The 1000 minima of the sample paths are summarised in Fig. \ref{img:event82}. Additionally, we estimate the interquartile ranges of the timing of the event from GISP2 to be (8.12, 8.16, 8.18) and GRIP to be (8.12, 8.14, 8.18) k cal yr BP. Our findings are consistent with previous studies reported elsewhere; but no quantification of the uncertainty has previously been attempted; see, for instance, \citet{thomas20078, kobashi2007precise}. 

An ensemble of sample paths thus provide a flexible data product in its own right. Indeed in climate reconstruction, this is precisely as proposed by \citet{tingley2012piecing}. The use above for minima is illustrative; any function of the process may be studied, conditional on the data.  One function, the conditional mean at time $t_g$, is of course more efficiently computed from its analytical expression as discussed in Eq. \eqref{eq:cGmean}.

\section{Alternative models}\label{sec:comparisons}
In this section, we discuss the model choice for $\mathbf\Sigma$ within the context of our application by comparing it with alternative models and choosing the best among them. To formally measure the benefit of joint modelling we compare it with an approach that ignores cross-correlation between cores; this was informally introduced in section \ref{sec:results}. Then, we propose an extension to the current model which assumes varying variances for different cores.

Let $M_1$ denote the structure for $\mathbf\Sigma$ defined in Eq. \eqref{eq:SigmaM1}. Recall that this assumes equal variance of increments across the cores. We will compare its performance with alternative models, $M_2$ and $M_3$. Model $M_2$ ignore the cross-correlation relationship between cores; as discussed in Sect. \ref{sec:results}. Hence it effectively comprises of two separate models; one for each core. We have demonstrated that the joint approach utilises information from both cores more effectively than the separate approach. We now define $M_3$ to have an additional parameter in the covariance structure to allow for varying variance between cores, i.e. 
\begin{equation}\label{eq:SigmaM3}
\boldsymbol\Sigma_a = 
\begin{pmatrix}
	1 & \rho\sqrt{a} \\
	\rho\sqrt{a} & a
	\end{pmatrix}
\end{equation}

The marginal posterior distribution for parameter $a$ of model $M_3$ is presented Fig. \ref{img:hyperPost4}. The mode centres around 1 while the marginal posteriors for other parameters (not shown here) are practically the same as those in model $M_1$ presented in Fig. \ref{fig:hyper}. These suggest that the common variance assumption of model $M_1$ is most suitable for this data set. Thus, model $M_3$ is essentially a more conservative version of $M_1$, but such an extension may not be necessary when we take into account the additional computational cost associated with an extra hyperparameter. Finally, Table \ref{tab:bic} provides further evidence that the extra parameter can be made redundant, as the Bayesian information criterion (BIC) of \citet{schwarz1978estimating} suggests that $M_1$ is superior to $M_3$.

\conclusions\label{sec:discussion}  

We have presented a hierarchical Bayesian model to jointly analyse multiple misaligned time series. An important component of our model is the Gaussian Markov assumption based on multivariate independent increments that gives us a natural vehicle for joint modelling. We further derived and implemented a fast algorithm for parameter inference based on this model. We applied our method to a pair of GISP2 and GRIP cores, and created data products that are consistent with our empirical knowledge of the physical climate system. More generally, we demonstrated that the joint approach utilises information from multiple cores more efficiently than one-core-at-a-time alternatives.

To the best of our knowledge, our work is a first attempt at directly addressing the joint behaviour of multiple ice cores in their raw and misaligned form.  We offer an approach to create data product in the form of non-Gaussian posterior predictive distributions, which is more flexible than what was previously possible. Additionally, our path sampling formulation allows for studies of a variety of non-linear functionals of partially observed processes. Here it is available at relatively little cost.


Some parameters in our model were formulated according to the respective length of the sections of ice cores. Since we consider the fairly stable period of climate conditions, we have not modelled the non-linear age-depth relationship. Furthermore, we did not take into account the error in the dating. Both of these assumptions are likely to be problematic in studying longer sections than the Holocene. A more realistic approach then is to allow for varying change of support. Within the hierarchical Bayesian framework, it is also conceptually straight forward to incorporate uncertainty in the time scale. 


In summary, this paper has been tailored to creating data products from Greenland ice cores. In this respect, we feel that it is a proof-of-concept building on the well-established ground of space-time modelling. There are several ways in which our model can be immediately extended to a wide class of spatio-temporal process. For example, we might investigate the possibility of including heterogeneous variability with respect to time by modifying Eq. \eqref{eq:state}. This corresponds to the spatial extension of the stochastic volatility model for irregular time series recently discussed by \citet{parnell2014bayesian}.


\appendix
\section{Implication of the change of support theory in the measurement procedure}\label{app:cosp}

In this appendix we discuss our treatment for the process variance and nugget via the change of support theory. More specifically, we examine the change in the theoretical semivariogram when there is a change in the underlying support of the data. Without loss of generality, we suppose that the data process $\boldsymbol{y}$ can be modelled by a noisy univariate independent increments process with a theoretical linear semivariogram of the form $\hat{\sigma}^2 + \frac{1}{2}\hat{v}^2\left|h\right|$ where the nugget $\hat{\sigma}^2$ is the intercept, the process variance $\hat{v}^2$ is twice the slope value, and $h$ is the time lag. We create the new process $\tilde{\boldsymbol{y}}$ on a new support by averaging $\boldsymbol{y}$ at every non-overlapping $w$ 'windows' of time. We are interested in the relationship between the semivariogram of $\boldsymbol{y}$ and that of $\tilde{\boldsymbol{y}}$. 


If $\boldsymbol{y}$ is a pure nugget process it can be shown that the semivariogram of the averaged process is $\frac{\hat{\sigma}^2}{w}$ when $ |h| \ge w$ and $\frac{\hat{\sigma} |h|}{w^2}$ when $|h| < w$. For the independent increments case, this is $\frac{1}{2}\hat{v}^2 |h| - \frac{w\hat{v}^2}{6}$ when $|h| \ge w$ and $\frac{\hat{v}^2 {|h|}^2}{2w} - \frac{\hat{v}^2 {|h|}^3}{6w^2} )$ when $ |h| < w$. We refer to \citet[Chap. 2.4]{chiles2012geostatistics} for the technical details of the aforementioned results. The only lags available for semivariogram calculation in practice are $|h| \ge w$. Thus, for a process that has both the nugget and independent increments, the implication is twofold. First, $\boldsymbol{y}$ and $\tilde{\boldsymbol{y}}$ share the same process variance $\hat{v}^2$. Second, their nugget parameters are (approximately) directly proportional. This approximation affects the semivariogram near the origin, where we use the linear function to account for the true cubic. From the analytical expression it can be see that the accuracy of the approximation depends on the value of $w$, and the relative difference between $\hat{\sigma}^2$ and $\hat{v}^2$.

For our application, we assume that the nugget effect is at an annual level (denoted as $\sigma^2_{annual}$),  with corresponding $\sigma^2_{GRIP} = \frac{\sigma^2_{annual}}{w_{GRIP}}$ and $\sigma^2_{GISP2} = \frac{\sigma^2_{annual}}{w_{GISP2}}$, where factors $w$'s denote the time support for each series. We can reparameterise in terms of $\sigma^2_\epsilon = \sigma^2_{GRIP}$, such that $\sigma^2_{GISP} = \sigma^2_\epsilon \times \frac{w_{GRIP}}{w_{GISP}} = \sigma^2_\epsilon\times k$. We set $\frac{55}{200}$ or 0.275 as the value for $k$, corresponding to their respective length of support. This value is also consistent with the descriptive statistics of the age increments, as represented in Fig. \ref{fig:input1}\textbf{(c)}. The sections, being 55cm and 200cm in lengths respectively for GRIP and GISP2, are negligible compare to the total length of the ice core which is about 3km (approximately 1.6 km of which covers the Holocene period). Thus we feel this is a reasonable approximation. 

\section{Model validation}\label{app:cv}
%


As a final model checking step, we determine whether the parameters in our model are identifiable or not. We do this by simulating model parameters ($v^2,\rho,\sigma^2_\epsilon$) based on the results of the data analysis of Greenland ice core, thence the latent bivariate process $\boldsymbol{x}$ and consequently artificial data $\boldsymbol{y}$. We partially average the sequences so as to match the change of support that occurs in our ice core example. We then fit the model (as described in Sect. \ref{sec:postHyper}) and determine whether the 50 and 90\% posterior intervals contain the true values. We repeat these steps 1000 times, and count up the proportion of occurrences where the intervals contain the true values. A properly calibrated and identifiable 50\% interval should contain the true value 50\% of the time, and similarly with the 90\% interval.

The results of our simulations are shown in Table \ref{tab:cv3}. As can be seen, the intervals contain slightly fewer than the desired proportion of true values, so our posterior intervals are over-precise. However, this effect appears small, and the model seems generally identifiable.

\begin{acknowledgements}
The authors thank professor Eric W. Wolff for his helpful comments on the Greenland ice core dataset.
\end{acknowledgements}



%
%

 \bibliographystyle{copernicus}
 \bibliography{/home/thinh/Documents/Learning/Dropbox/PhD/Spatial_Stat/STModels/BivariateModel/CASIpaper/biblio.bib}



\addtocounter{table}{0}\renewcommand{\thetable}{\arabic{table}}

%
%
%
%
%
%

\begin{table*}[th]
\begin{center}
\begin{tabular}{lcccc}
\hline
Model & Covariance structure for two cores & $-2\log{L}$ & Penalty & BIC\\
\hline
\hline
$~~M_1$ & \( \begin{pmatrix}
	1 & \rho \\
	\rho & 1
	\end{pmatrix} \) & 44998 &  25 &  \underline{45023} \\
$~~M_3$ & \( \begin{pmatrix}
	1 & \rho\sqrt{a} \\
	\rho\sqrt{a} & a
	\end{pmatrix} \) & 44998 & 33 & 45031 \\
\hline
\end{tabular}
\end{center}
\caption{Values of Bayesian Information Criterion (BIC) obtained from different model setting. The underlined value highlights the model with the best fit.}
\label{tab:bic}
\end{table*}

\begin{table*}
\begin{center}
\begin{tabular}{lccccc}
\hline
&  & Sample 1 & Sample 2 & $\hdots$ & \\
\hline
\hline
Hyperparameters & & (0.2, 0.8, 0.4) & (0.3, 0.9, 0.5) & $\hdots$ \\
\hline
Time & 7.5 & -34.3 & -34.7 & $\hdots$ \\ 

 & 8 & -35.0 & -34.2 & $\hdots$ \\ 

 & 8.5 & -34.4 & -34.4 & $\hdots$ \\ 

 & 9 & -34.6 & -34.2 & $\hdots$ \\ 
\hline
\hline
Minimum &  & -35.0 & -34.7 & $\hdots$ \\ 
Time of minimum &  & 8 & 7.5 & $\hdots$ \\ 
\end{tabular}
\end{center}
\caption{Illustration of sample paths from one core on a time grid of 500 years intervals. For each sample, we simulate the hyperparameters from their joint posterior distribution, followed by the sample paths of the latent process $\boldsymbol{x_g}$. The non-linear functionals of interest are displayed in the last two rows.}
\label{tab:82ka}
\end{table*}

\addtocounter{table}{-2}\renewcommand{\thetable}{B\arabic{table}}

\begin{table*}
\begin{center}
\begin{tabular}{lcc}
\hline
Parameter & Proportion inside 50\% CI & Proportion inside 90\% CI \\
\hline
\hline
$\boldsymbol{x}$ & 51\% & 89\% \\
$v^2$ &  48\% & 88\% \\
$\rho$ & 49\% & 90\% \\
$\sigma^2_\epsilon$ & 48\% & 90\% \\
\hline
\end{tabular}
\end{center}
\caption{Performance of the model fitting algorithm. All results were based on 1000 simulation runs.}
\label{tab:cv3}
\end{table*}

\addtocounter{figure}{0}\renewcommand{\thefigure}{\arabic{figure}}

\begin{figure}[h]
\includegraphics[width=8.3cm]{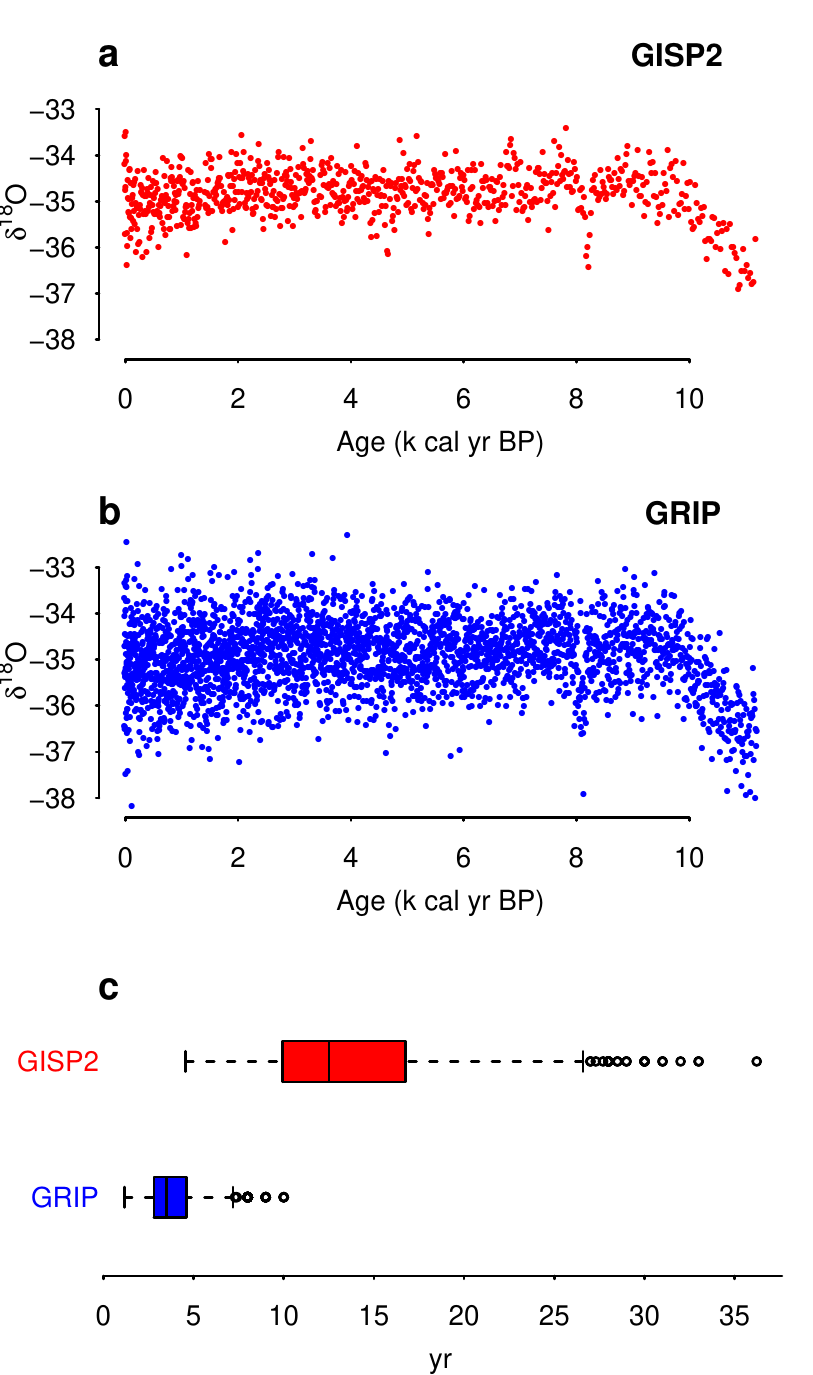}
\caption{Scatter plots of $\delta^{18}O$ measurements and ages of \textbf{(a)} GISP2 and \textbf{(b)} GRIP. \textbf{(c)} Boxplots of the time increments clearly show different irregularities in the ages. The boxplot for GISP excludes an age difference value of 80.6 yr between roughly 1.32 and 1.4 k cal yr BP. The interquartile range (calculation including the omitted value) are (10.0, 12.5, 16.8) and (2.8, 3.5, 4.6) year for GISP2 and  GRIP respectively.}
\label{fig:input1}
\end{figure}

\begin{figure}
\includegraphics[width=8.3cm]{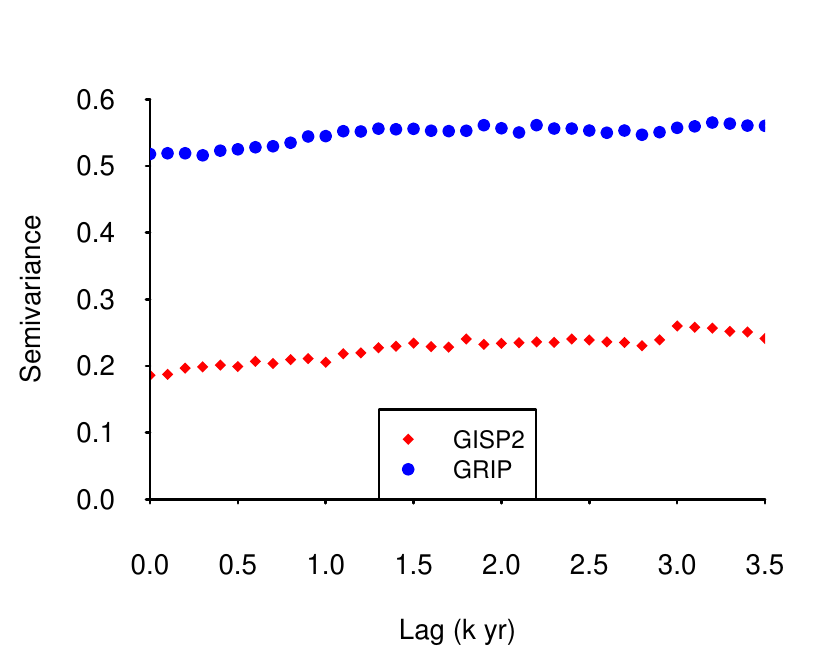}
\caption{Empirical semivariograms of GISP2 and GRIP. They suggest that the linear variogram is a suitable model for both of the ice core data sets.}
\label{fig:cosemp}
\end{figure}

\begin{figure*}
\includegraphics[width=14cm]{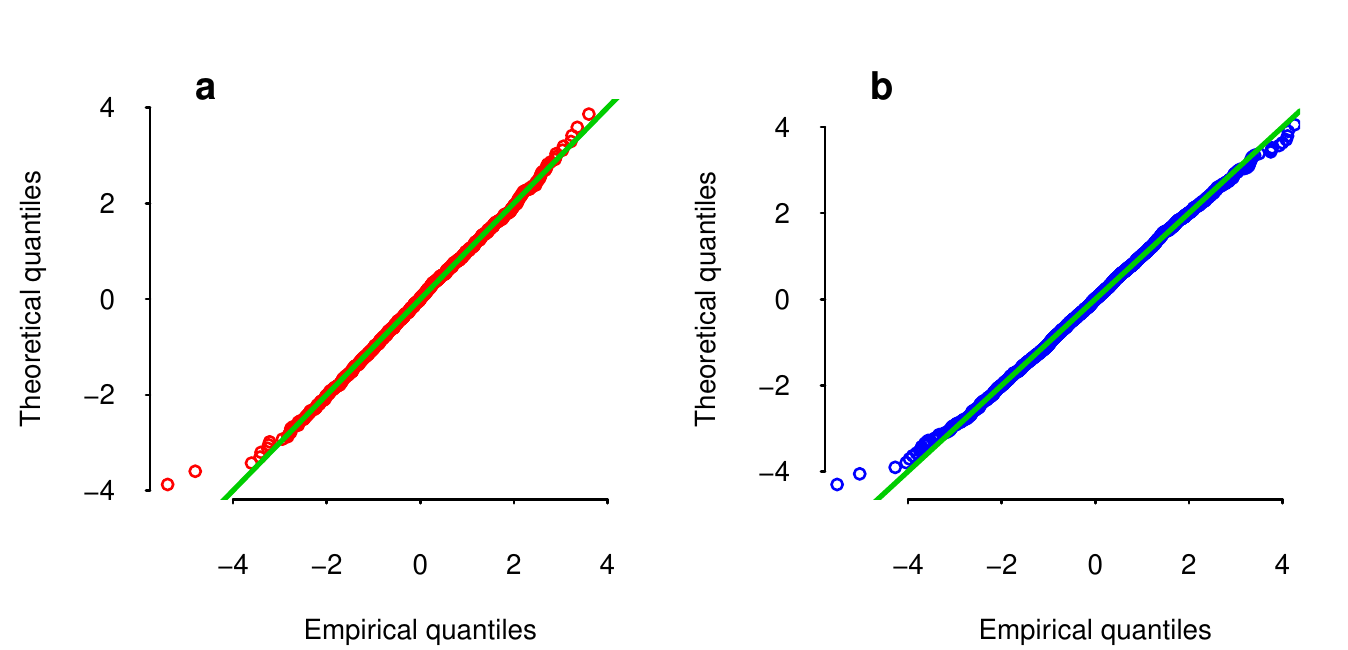}
\caption{QQ plots of the standardised increments, i.e. the ratio of the first differences in the $\delta^{18}O$ measurements and estimated standard errors of increments for \textbf{(a)} GISP2 and \textbf{(b)} GRIP. The unusual values correspond to consecutive pairs of measurements with very large difference in the $\delta^{18}O$ values.} 
\label{fig:input2}
\end{figure*}

\begin{figure*}
\includegraphics[width=13cm]{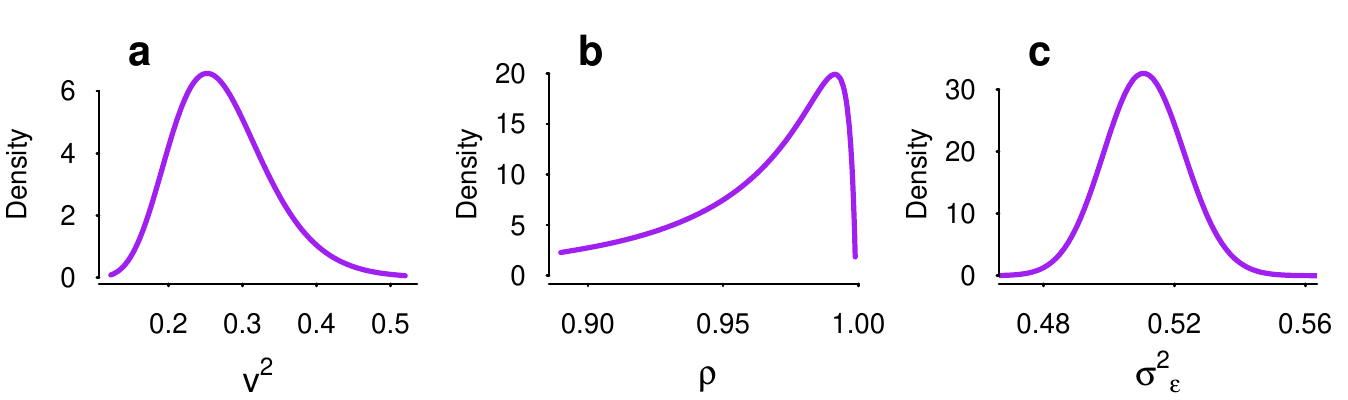}
\caption{Plots of the smoothed marginal posterior distributions of \textbf{(a)} the variance of the unit increment of $\boldsymbol{x}$, \textbf{(b)} cross correlation and \textbf{(c)} nugget parameter for GRIP. These are results based on the joint model discussed in Sect. \ref{sec:fullmodels} applied to both GISP2 and GRIP.}
\label{fig:hyper}
\end{figure*}

 \begin{figure*}
      \includegraphics[width=14cm]{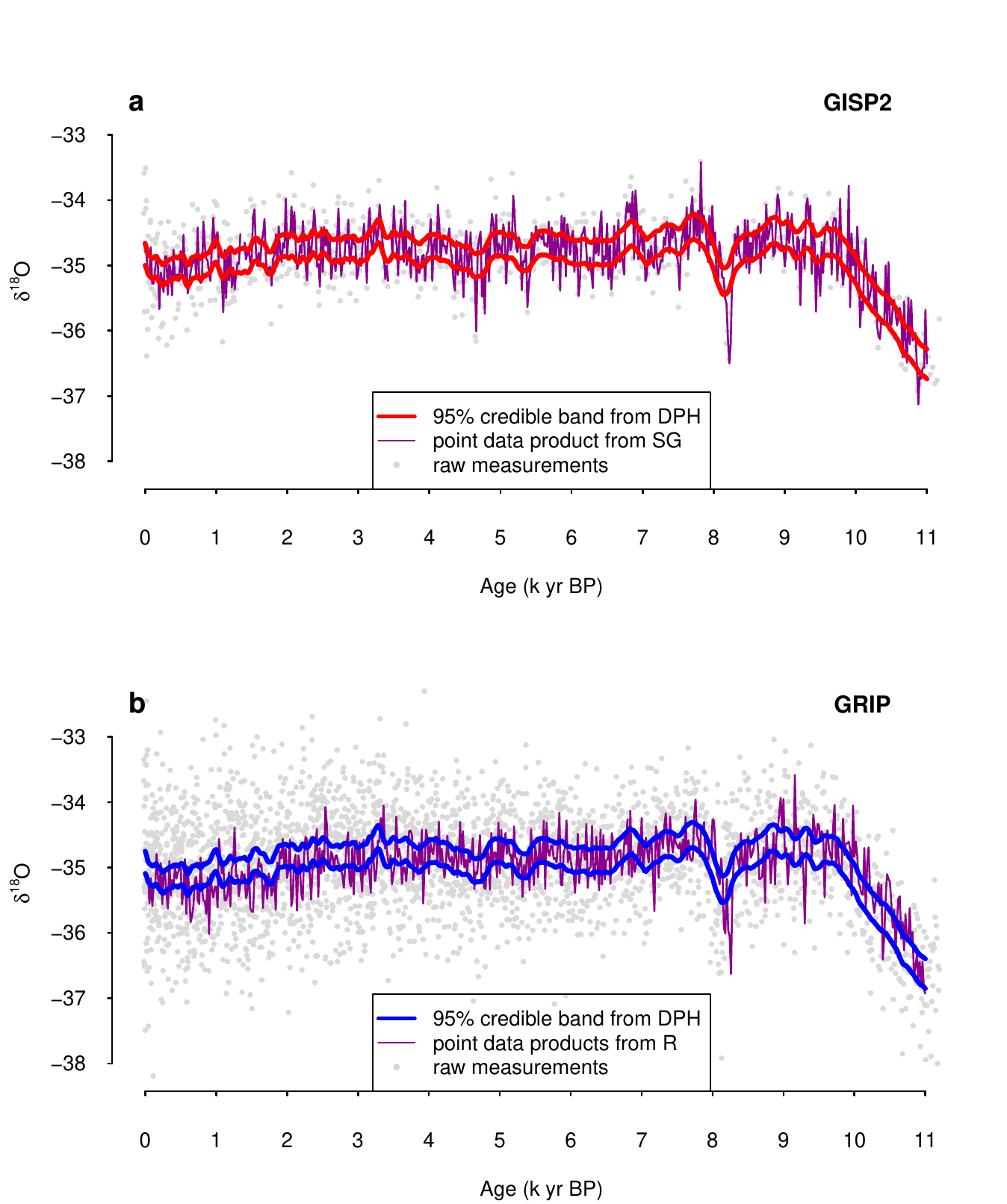}
      \caption{Plots of quantile-based 95\% credible intervals of the marginal posterior distributions of process $\boldsymbol{x_g}$ on a bidecadal time grid over the period of 0 to 11k cal BP conditional on both GISP2 and GRIP. We also show the bidecadal data product from \citet[SG]{stuiver2000gisp2} and \citet[R]{rasmussen2006new}.}
		\label{fig:cG}
    \end{figure*}


 \begin{figure*}
      \includegraphics[width=14cm]{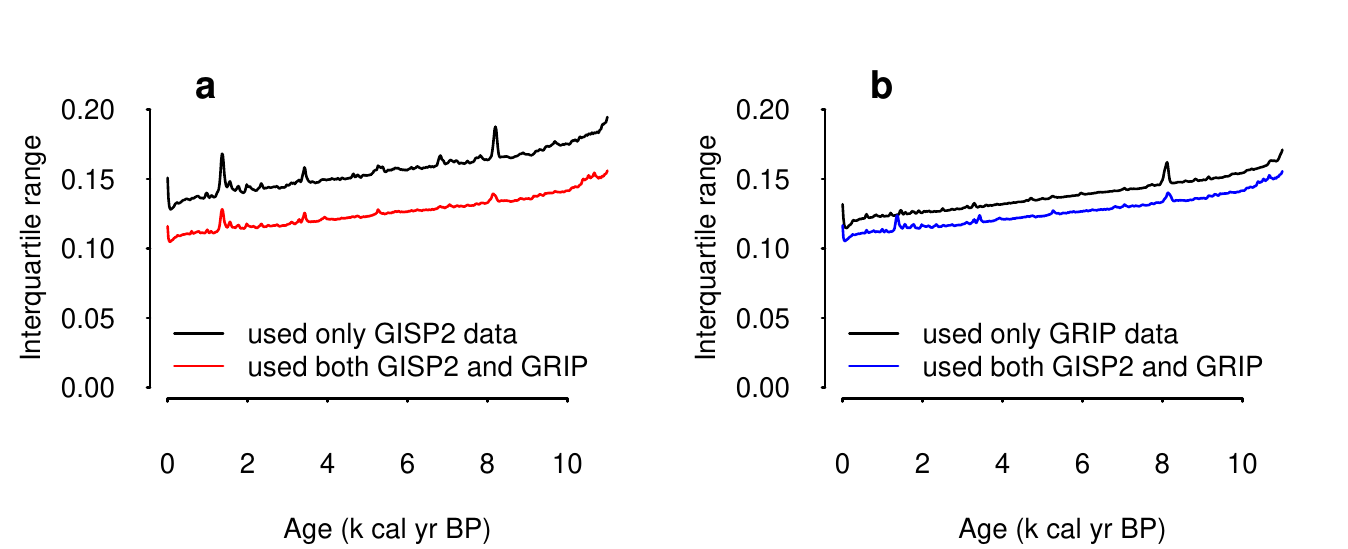}
      \caption{Plots of the interquartile ranges of the marginal posterior distributions of elements of $\boldsymbol{x_g}$ on a bidecadal time grid over the period of 0 to 11k cal BP at core \textbf{(a)} GISP2 and \textbf{(b)} GRIP. The main features from these plots are: (i) interquartile ranges of the separate models are always higher than those of the joint model; (ii) the gaps of the differences in \textbf{(a)} are consistently larger than those in \textbf{(b)}; (iii) there are several spikes; (iv) a slight tendency for increased IQR further back in time. See the text for a detailed discussion.}
		\label{img:indep}
    \end{figure*}

\begin{figure*}
\includegraphics[width=14cm]{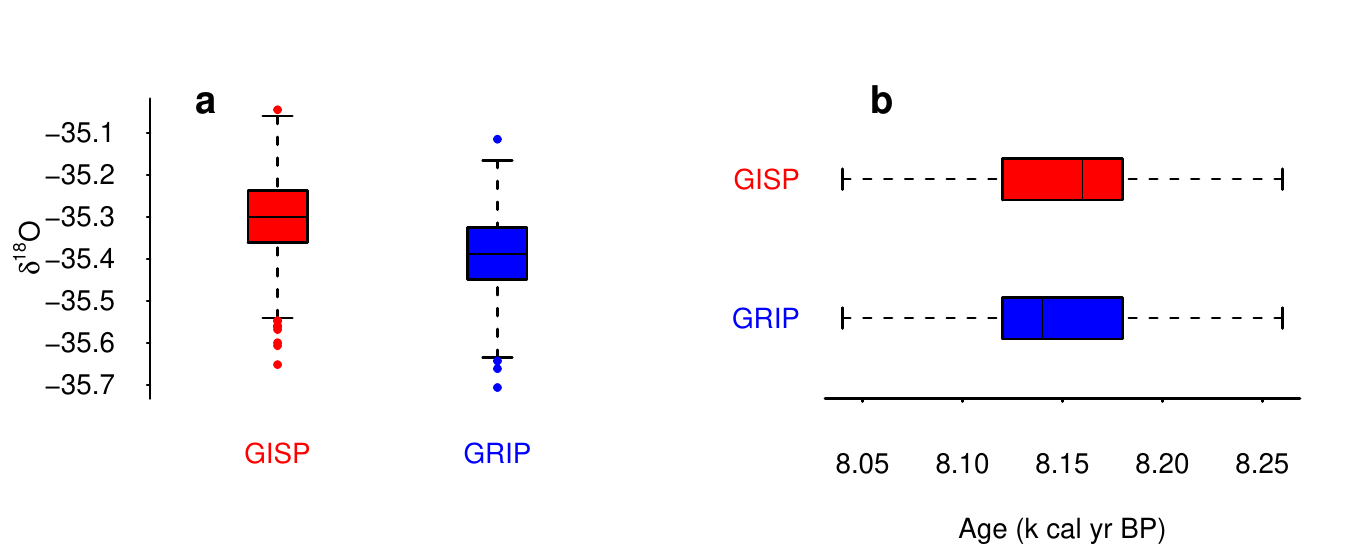}
\caption{\textbf{(a)} Boxplots of the minima and \textbf{(b)} time of the minima from GISP2 and GRIP over the range of 7.9 to 8.5 k cal yr BP. The interquartile range of the timing of the event from GISP2 are (8.12, 8.16, 8.18) k cal yr BP, and this is (8.12, 8.14, 8.18) for GRIP. All estimates are based on 1000 sample paths.}
\label{img:event82}
\end{figure*}

    \begin{figure}
      \includegraphics[width=8.3cm]{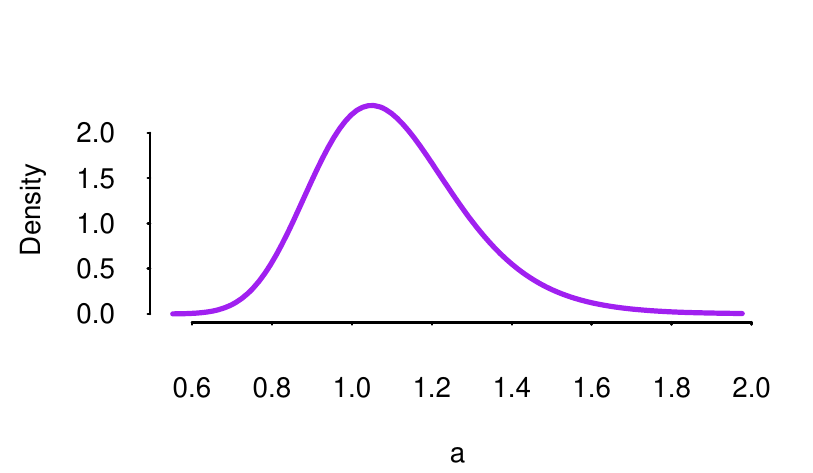}
      \caption{Plot of the smoothed marginal posterior distribution of the parameter $a$ in the covariance matrix of model $M_3$ described in Sect. \ref{sec:comparisons}. The mode is roughly 1 which suggests that $a$ is redundant.}
		\label{img:hyperPost4}
    \end{figure}

\end{document}